\documentclass[fleqn,usenatbib]{mnras}

\usepackage[T1]{fontenc}

\usepackage{natbib}
\usepackage{graphicx}	
\usepackage{amssymb}	
\usepackage{textcomp}
\usepackage{comment}
\usepackage[utf8]{inputenc}
\usepackage{multirow}
\usepackage{appendix}
\usepackage{float}
\usepackage{gensymb}
\usepackage{setspace}

\title[Earth-Like Stratospheric Clouds with JWST]{Stratospheric Clouds Do Not Impede JWST Transit Spectroscopy for Exoplanets with Earth-Like Atmospheres}

\author[Doshi, Cowan \& Huang]{
Dhvani Doshi,$^{1,2}$\thanks{Email: dccdoshi@uwaterloo.ca} 
Nicolas B. Cowan$^{2,3,4,5}$\thanks{Email: nicolas.cowan@mcgill.ca}
\& Yi Huang$^{5,6}$\thanks{Email: yi.huang@mcgill.ca}
\\
$^{1}$University of Waterloo, 200 University Ave W, Waterloo, ON N2L 3G1, Canada\\
$^{2}$Institut de recherche sur les exoplanètes, Université de Montréal, C.P. 6128, Succ. Centre-ville, Montréal, QC H3C 3J7, Canada\\
$^{3}$Department of Earth \& Planetary Sciences, McGill University, 3450 rue University, Montréal, QC H3A 0E8, Canada\\
$^{4}$Department of Physics, McGill University, 3600 rue University, Montréal, QC H3A 2T8, Canada\\
$^{5}$McGill Space Institute, 3550 rue University, Montréal, QC H3A 2A7, Canada\\
$^{6}$Department of Atmospheric \& Oceanic Sciences, McGill University, 805 Sherbrooke St.\ West, Montréal, Québec H3A 0B9, Canada
}

\date{Accepted XXX. Received YYY; in original form ZZZ}

\pubyear{2022}

\begin{document}
\label{firstpage}
\pagerange{\pageref{firstpage}--\pageref{lastpage}}
\maketitle

\begin{abstract}
The James Webb Space Telescope (JWST) will provide an opportunity to investigate the atmospheres of potentially habitable planets. Aerosols, significantly mute molecular features in transit spectra because they prevent light from probing the deeper layers of the atmosphere. Earth occasionally has stratospheric/high tropospheric clouds at 15-20 km that could substantially limit the observable depth of the underlying atmosphere. We use solar occultations of Earth’s atmosphere to create synthetic JWST transit spectra of Earth analogs orbiting dwarf stars. Unlike previous investigations, we consider both clear and cloudy sightlines from the SCISAT satellite. We find that the maximum difference in effective thickness of the atmosphere between a clear and globally cloudy atmosphere is 8.5 km at 2.28 microns with a resolution of 0.02 microns. After incorporating the effects of refraction and Pandexo’s noise modelling, we find that JWST would not be able to detect Earth like stratospheric clouds if an exo-Earth was present in the TRAPPIST-1 system, as the cloud spectrum differs from the clear spectrum by a maximum of 10 ppm. These stratospheric clouds are also not robustly detected by TauREx when performing spectral retrieval for a cloudy TRAPPIST-1 planet. However, if an Earth size planet were to orbit in a white dwarf’s habitable zone, then, we predict that JWST's NIRSpec would be able to detect its stratospheric clouds after only 4 transits. We conclude that stratospheric clouds would not impede JWST transit spectroscopy or the detection of biosignatures for Earth-like atmospheres.
\end{abstract}



\begingroup
\let\clearpage\relax
\endgroup
\newpage

\section{Introduction}
When an extrasolar planet transits in front of its star, some of the starlight reaching a distant observer is filtered through the atmosphere of the planet. By comparing the spectrum of the star during such a planetary transit to its spectrum at other times, one can obtain a transmittance spectrum of the planet's atmosphere~\citep{SeagerSasselov2000}.  Transit spectroscopy is currently the most prolific technique for determining the composition of exoplanet atmospheres, which provides insights into their formation, evolution and climate \citep{Kreidberg2018,doi:10.1146/annurev-astro-081817-051846}. 

The atmospheric composition of planets orbiting in the habitable zone of their host stars is of particular interest. If these planets are Earth-like in other respects, then they should be able to harbour liquid water at their surface and hence life as we know it~\citep{Kasting1993}.  Since the trace gases in Earth's atmosphere are symptomatic of our biosphere~\citep{Sagan1993}, there is hope that next generation telescopes could detect such atmospheric biosignatures on temperate terrestrial exoplanets \citep{2002AsBio...2..153D}. Genuine Earth twin transits are unlikely and would occur infrequently, so transit spectroscopy is only feasible in the near term for planets orbiting red dwarf stars, and even then they will be daunting \citep{2015PASP..127..311C,2016MNRAS.461L..92B,Morley2017,Krissansen-Totton2018,2019AJ....158...27L,Macdonald,Evans2021}

\subsection{The Impact of Aerosols on Transit Spectroscopy}
In addition to the small signal, transit spectroscopy of exoplanets is made more difficult by the presence of aerosols \citep{2021A&G....62.1.36B}.  Be they photochemically produced hazes or condensate clouds, these small particles tend to scatter and absorb radiation over a wide range of wavelengths, hence obscuring the spectral features due to atoms and molecules \citep{2014PNAS..11112601B,2020SSRv..216...82B}. Roughly speaking, aerosols present at some height in the atmosphere of an exoplanet make it difficult to probe deeper layers in the atmosphere \citep{2014Natur.505...69K}.

Even in the absence of clouds, the deeper layers of an atmosphere---those close to the surface---are hard to probe with transit spectroscopy.  Since an atmosphere is densest at the bottom, it is liable to completely block all light, regardless of wavelength \citep{Kalt&Traub}.  Moreover, depending on the angular size of the host star as seen from the exoplanet, refraction of light tends to bend light out of the line of sight of a distant observer \citep{2010ApJ...720..904S,2014ApJ...791....7B,2017ApJ...850..128R}. As a result of these effects, the lower atmosphere of most exoplanets is inscrutable via transit spectroscopy \citep{2018ApJ...865...12B}.

While, photochemically produced hazes are expected to overpower the effects of condensate clouds for rocky exoplanets such as TRAPPIST-1e~\citep{Fauchez2019} and the Archean Earth atmosphere which likely harboured of an organic haze~\citep{Coustenis1995,Clarke1997}, hazes do not interfere with the spectroscopy for modern Earth-like atmospheres. Therefore, for a habitable planet like the Earth, the most important aerosols are H$_2$O clouds, which are usually limited to the lower atmosphere, where temperature and moisture are greater. But those regions of the atmosphere are nearly impossible to probe via transit spectroscopy in any case. To first order, therefore, one does not expect clouds to pose a challenge to transit spectroscopy of exoplanets with an Earth-like atmosphere. 
\subsection{Stratospheric Clouds on Earth}
More than 99\% of the water in Earth's atmosphere is concentrated in the troposphere. In contrast, the stratosphere is dry, with the volume mixing ratio of water vapour being typically several parts-per-million, which makes it hard for H$_2$O clouds to form there. Nevertheless, clouds are observed in the stratosphere. Meteorologists have kept century-long records of stratospheric clouds \citep{stanford1974century}. They are preferentially observed in the winter-time polar stratosphere, due to the extreme cold temperatures in the polar vortices \citep{salby1996fundamentals}. They can also be generated by atmospheric gravity waves \citep{dornbrack2002evidence} or be detrained from the overshooting convection towers \citep{wang2013physics}.

Stratospheric clouds are of great interest to climatologists because they are particularly sensitive to climate change \citep{wetherald1986investigation}. Modeling assessments showed that a small change in stratospheric optical depth due to stratospheric-cloudiness can lead to significant impacts on the Earth's radiation budget and thus modify the extent of climate warming \citep{harshvardhan1979perturbation,ramanathan1988greenhouse}. 

Tropical cirrus clouds exhibit a range of physical appearances from wide sheets to wispy filaments. Consequently there is a wide variability in particle size and number density but generally cirrus clouds are composed of non-spherical ice particles. They are optically thin, but absorb and re-emit infrared radiation from the Earth. Cirrus clouds cover up to 30\% of Earth's surface and thus may contribute to global warming as their relatively cold cloud-top temperatures reduce the outgoing long wave radiation to space relative to an equivalent cloud-free region.~\citep{Lynch1996,Zondolo}. 

Polar stratospheric clouds are rare and appear in the winter polar stratosphere. Although the stratosphere is already very dry and cold, polar stratospheric clouds require even lower temperature, close to -80 \degree C. They form at similar altitudes as the ozone layer and they facilitate chlorine depletion of ozone via heterogeneous chemistry \citep{solomon1990progress}. Polar stratospheric clouds hence are considered to be partially responsible for the ozone holes over the polar regions~\citep{Tritscher2021}.

\subsection{Cloud Modelling}
General circulation models (GCMs) of synchronously-rotating planets orbiting M-dwarfs have been used to predict the location and optical properties of H$_2$O clouds and hence their effect on transit spectroscopy~\citep{Fujii2017,Komacek,Pidhorodetska2020,Suissa,May2021,Mikal-Evans,Ding2022}. Inter-model comparisons suggest that differences in cloud parametrization lead to a $\sim$40\% systematic uncertainty \citep{Fauchez2021}, making it difficult to ascertain the impact of high-altitude clouds on transit spectra. More importantly, the dearth of empirical constraints means that we still do not know whether M-Earths have atmospheres, let alone whether they match model predictions. As a result, developing an empirical transit spectrum from real data, such as solar occultations, can offer complementary insights into the impact of high altitude clouds, even if M-Earths are unlikely to have an Earth-like atmosphere due to the redder incident spectrum and likely synchronous rotation of the planet. While an Earth-based spectrum limits our understanding to one type of atmosphere, it can benchmark our expectations as to how Earth-like atmospheric conditions may impact transit spectroscopy on other planetary systems~\citep{Robinson2018}.
\subsection{Outline of Paper}
Solar occultation data from the SCISAT satellite have already been used to assess Earth's transit spectrum in the absence of clouds~\citep{Schreier2018,Macdonald}. In this paper, we set out to estimate the impact of high-altitude clouds on transit spectroscopy using Earth observations from the SCISAT ACE-FTS instrument. In Section~\ref{data} we explain the different data and models to create the effective thickness and transit spectra in Section~\ref{transit}. We apply this Earth-like atmosphere to different exoplanetary systems in Section~\ref{exoplanet}, and quantify the differences between cloudy and clear atmospheres in Section~\ref{characterization}. Finally, we discuss and summarize our findings in Section~\ref{discussion}.

\section{Model Description}
\label{data}
\subsection{Solar Occultation Spectroscopy}

Measurements taken during solar occultations mimic those of transits due to similarities in observational geometry. In both cases, grazing sunlight passes through the planet's upper atmosphere. A solar occultation measurement is only sensitive to a single impact parameter while a transit spectrum simultaneously probes all impact parameters. We must therefore combine many solar occultation spectra corresponding to a range of impact parameters to simulate a transit spectrum.

We use solar occultation data from the Canadian satellite, SCISAT, to develop transit spectra from Earth's atmospheric properties. The primary instrument on SCISAT, Atmospheric Chemistry Experiment -- Fourier Transform Spectrometer (ACE-FTS), measures infrared atmospheric absorption signals during sunrise and sunset~\citep{Bernath}. To optimize global coverage, the SCISAT satellite operates on a high inclination ($75^{\circ}$) circular low Earth orbit ($640$ km). This allows for data collection from the tropics, mid-latitudes, and polar regions. ACE--FTS offers a sufficiently large signal-to-noise ratio (SNR) due to its highly folded double pass optical design. The instrument produces a vertical profile for Earth's atmospheric constituents by recording the atmospheric transmittance at a range of wavelengths at different altitudes. 

The vertical range of ACE--FTS is about $4\textup{--}128$~km, where the lower limit is dictated by the obstruction of low altitude clouds or the absence of the Sun from the instrument's line of sight. The exact altitudes at which the transmittance is measured within one occultation is governed by the beta angle, the angle between the satellite's orbital plane and the Earth-Sun vector. As a result, each occultation will sample the transmittance at a unique set of altitudes and have a different number of measurements. Multiple occultations can be stacked to further improve the SNR and provide a holistic image of Earth's atmosphere.

\subsection{Clear Atmosphere Data} 

We follow~\cite{Macdonald} to develop non-cloudy atmospheric transmittance spectra using the ACE--FTS Atlases~\citep{Hughes}. The longevity of the ACE mission has resulted in hundreds of occultations, thus the Atlases were created to provide a baseline for a non-cloudy atmosphere at high signal to noise ratio. They use data from occultations where clouds did not interfere with the spectra. Occultations were chosen based on latitude and season to create five different data-sets: ArcticWinter (60-90$\degree$N, Dec-Feb), ArcticSummer (60-90$\degree$N, Jun-Aug), MidLatWinter (30-60$\degree$N, Dec-Feb), MidLatSummer (30-60$\degree$N, Jun-Aug), and Tropics (30$\degree$N-30$\degree$S, Jan-Dec). 

The spectra from each occultation are divided into $4$~km bins in the range $4\textup{--}128$~km. This typically results in $800$ spectra within each bin, which are averaged to create one spectrum for each $4$~km bin. The transmittance in each spectrum is corrected between zero and unity, as calibrations errors  resulted in some transmittance data outside of this acceptable interval. The transmittance spectrum for a bin represents the transmittance at the bin's mid-point altitude. Therefore the bin of $4\textup{--}8$~km, contains a transmittance spectrum for an impact parameter of $6$ km. 

The ACE--FTS Atlases present transmittance spectra for wave numbers of $750\textup{--}4400$~cm$^{-1}$, but we focus on the range of $2.28\textup{--}13.32$~$\mu$m. The spectra are provided at a resolution of $0.0025$ cm$^{-1}$, but we bin them to $0.02$~$\mu$m. We construct a non-cloudy world-average by averaging the spectra of all five atlases.

\subsection{ACE Cloudy Data}
 
We create cloudy transmission spectra, using individual occultations from the ACE Mission that showed evidence of a stratospheric cloud. The scope of the cloudy transmission spectra is limited to stratospheric clouds because ACE--FTS stops measuring once it detects the presence of low altitude clouds, which have a higher aerosol extinction. The two types of clouds we study in this work are tropical cirrus clouds and polar stratospheric clouds. 

Occultations are binned and averaged following the ACE--FTS Atlases procedure for the tropical cirrus clouds and polar stratospheric clouds; we use four and two occultations respectively. The details of each occultation can be found in Appendix~\ref{A1}. The aerosol extinctions of the clouds present in these occultations match the average mean extinction values for stratospheric clouds~\citep{salby1996fundamentals}. Their high altitudes and typical extinctions allow them to represent the most cloudy scenario for transit spectroscopy of an Earth-like atmosphere.

\begin{figure}
    \centering
    \includegraphics[scale=0.47]{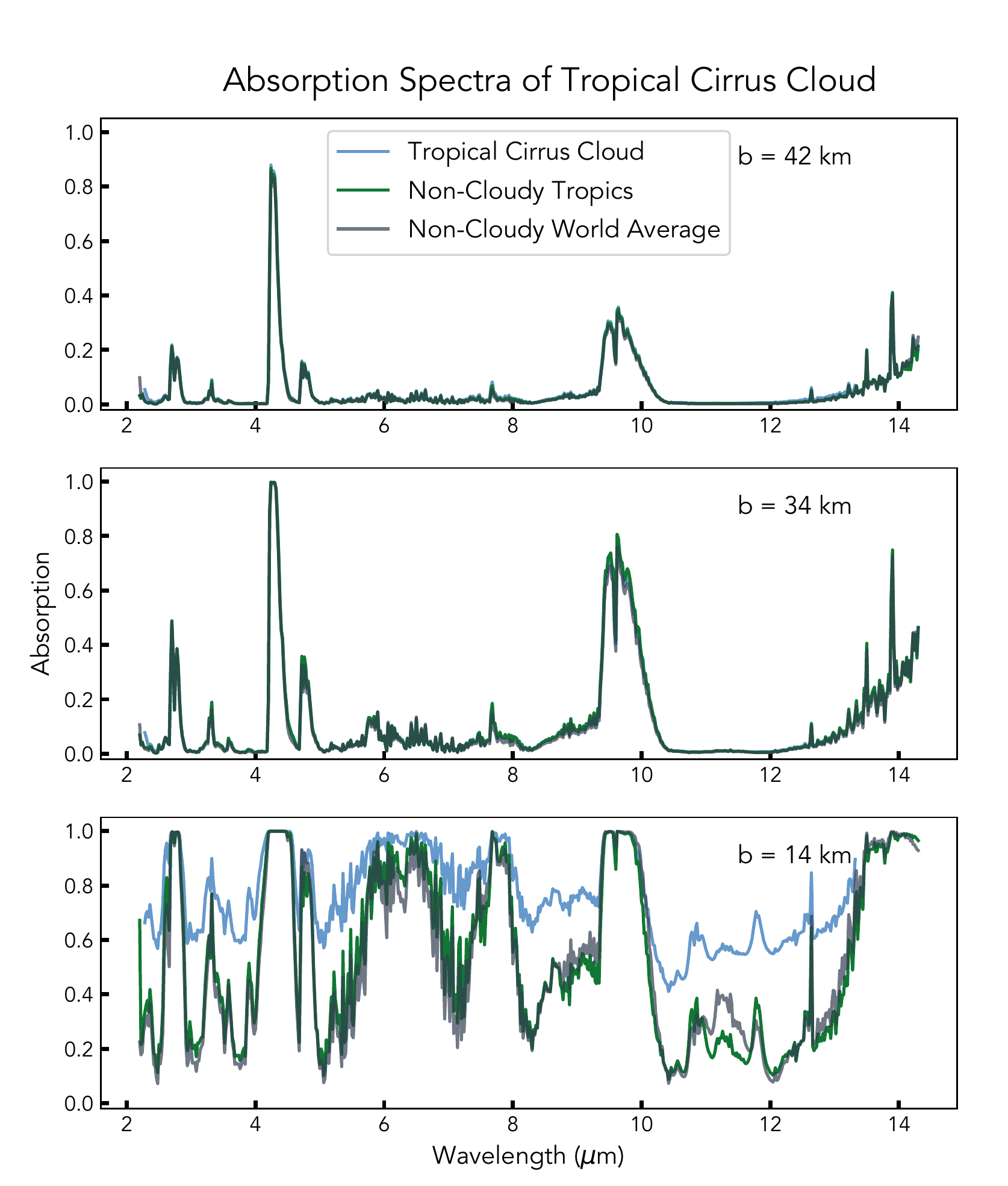}
    \includegraphics[scale=0.47]{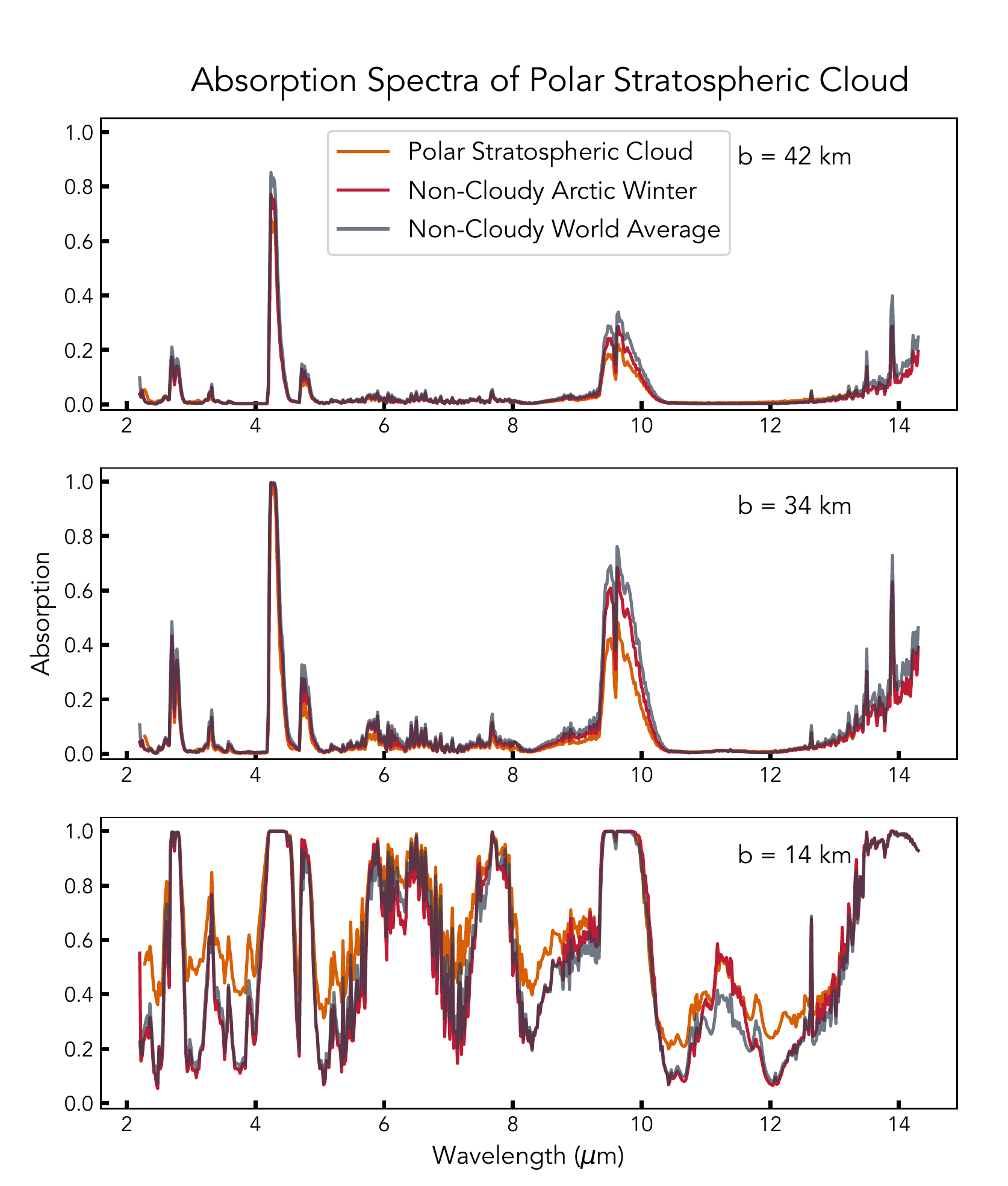}
    \caption{\emph{Top: }Transmittance for a tropical cirrus cloud, non-cloudy Tropics, and non-cloudy world average. The tropical cirrus cloud mainly affects the transmittance below the cloud deck. We can see slight deviations between the transmittance of the non-cloudy tropics and world average spectra, where the tropics present a lower transmittance. The tropical cirrus spectra match the non-cloudy tropics spectra better than the non-cloudy world average at higher altitudes. \emph{Bottom: }The polar stratospheric cloud transmittance deviates from the non-cloudy atmosphere at all altitudes above the cloud deck. While the non-cloudy arctic winter spectra also deviates from the non-cloudy world average, it still does not match the polar stratospheric spectra.}
    \label{fig:TransSpectra}
\end{figure}

We plot in Figure~\ref{fig:TransSpectra} the absorption spectra of scenes with and without stratospheric clouds at different altitudes above the Earth's surface. The tropical cirrus clouds mainly impact the absorption below their $\sim$17~km cloud deck and match the clear tropics atmosphere above the cloud deck. Meanwhile, the polar stratospheric clouds have an absorption spectrum that differs from the arctic winter and world average atmosphere at all altitudes.

\subsection{Synthetic Cloud Data}
In order to validate the ACE cloudy data, we create synthetic cloud transmittance observations. We empirically calculate the transmittance through a global cloud layer with a given cloud deck, thickness, and aerosol extinction using solar occultation geometry. The transmittance is calculated for the same impact parameters that are provided in the ACE--FTS Atlases. These synthetic cloud transmittance values are combined with the ACE world average clear observations to create synthetic solar occultation observations with a global cloud layer. In Figure~\ref{fig:EffH}, we show that synthetic clouds with a higher aerosol extinction than the clouds observed by ACE--FTS could strongly mute the effective thickness spectrum.

\section{Synthetic Transit Spectra}
\label{transit}
To create a synthetic transit spectrum from the ACE transmittance data, we follow the method of~\cite{Macdonald}, which was validated against the optical depth approximation. The transit depth, $D$, is related to the wavelength dependant effective height, $h_{\lambda}$, by~\citep{Brown}: 
\begin{equation}
    D_{\lambda} = \left(\frac{R_p + h_{\lambda}}{R_{*}}\right)^2,
    \label{eq:Deffheight}
\end{equation}
where $R_p$ and $R_*$ are the radii of the planet and star.

The transit depth can also be expressed in terms of the transmittance of the planet's atmosphere at various impact parameters, 

\begin{equation}
    D_{\lambda} = \left(\frac{R_p}{R_*}\right)^2 + \frac{2}{R_*^2}\int_{R_p}^{R_*}b\big(1-T(b,\lambda)\big)\,db,
    \label{eq:Dtrans}
\end{equation}
 where $b$ is the impact parameter, $T(b,\lambda)$ is the transmittance along the chord at the given impact parameter, and the upper limit is much greater than $R_p$ but not necessarily $R_*$.
 
Combining Equations~\ref{eq:Deffheight} and~\ref{eq:Dtrans}, we obtain the effective height of the atmosphere in terms of the ACE transmittance data, 
\begin{equation}
	h_{\lambda} = - R_{p} + \sqrt{{R_p}^{2}+2 \int_{R_p}^{R_p+b_{\rm max}} b\big(1-T(b,\lambda)\big)\,db} .
\label{eq:Effective_Thickness}
\end{equation}

We use Simpson's rule to approximate the integral in Equation~\ref{eq:Effective_Thickness}. The upper limit of the integral is taken as $R_p + 126$ km instead of $R_*$ because the highest impact parameter bin for ACE data is $b_{\rm{max}}=124\textup{--}128$~km. The transmittance converges to unity at impact parameters greater than $\sim$80~km for the full wavelength range. As a result, the transmittance behaviour above $126$~km does not change the effective height. 

However, the minimum ACE impact parameter of $6$~km is too high as a lower limit of the integral because the effective height is sensitive to the near surface transmittance behaviour. The optical depth is largely dictated by a decaying pressure exponential, therefore we fit a decaying exponential to the data and extrapolate the transmittance for impact parameters of $0$~km and $2$~km. This resolution was sufficient to provide an accurate measure of the effective height. 
   
\begin{figure*}
    \centering
    \includegraphics[scale=0.5]{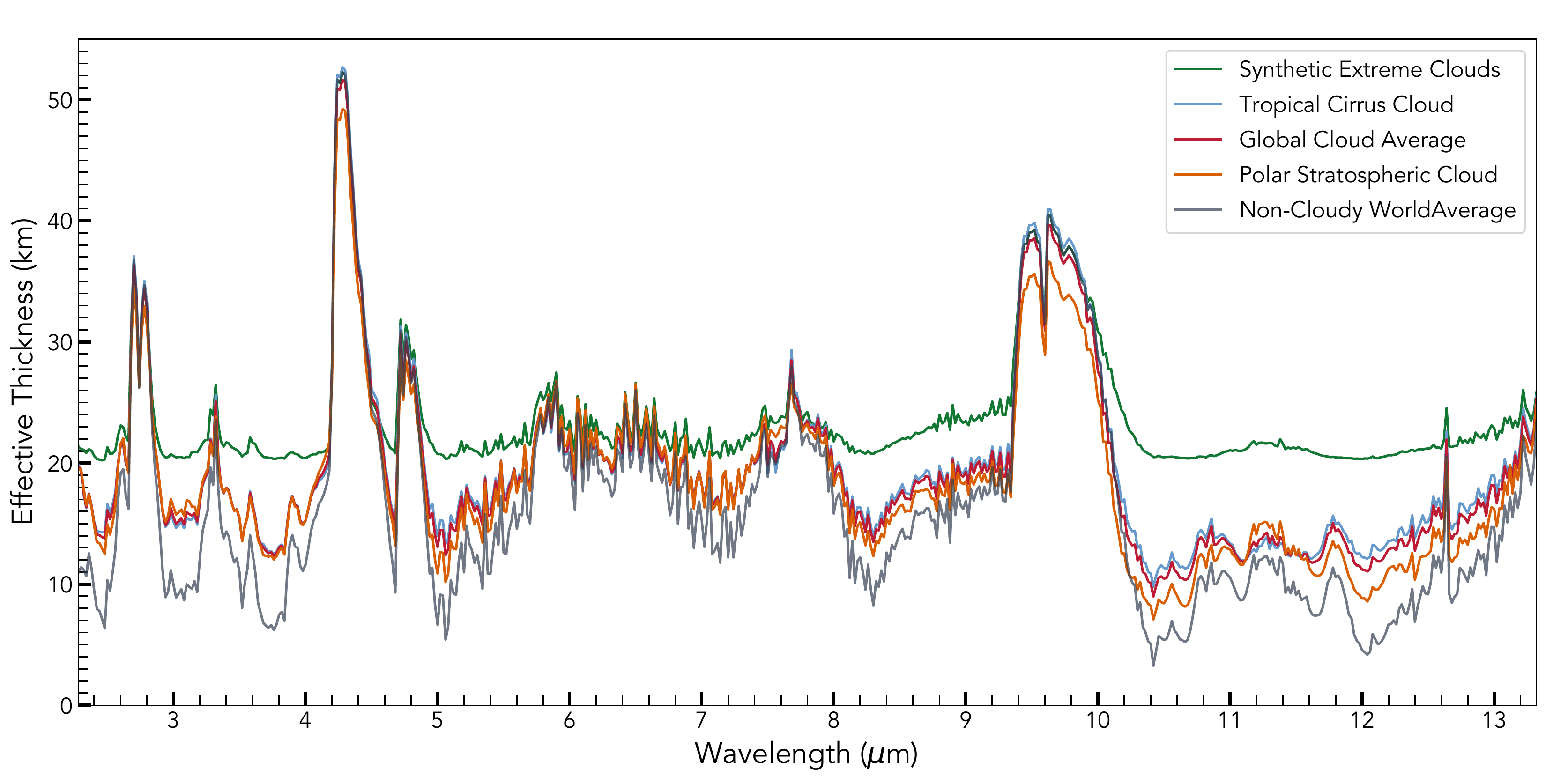}
    \caption{The effective thickness spectra for various Earth analog atmospheres. Clouds place a lower limit on how deep one can probe the atmosphere. The spectra for the polar stratospheric and tropical cirrus cloud data represent a global cloud layer with  atmospheric properties specific to that cloud type and location. The global cloud average represents an atmosphere with 70\% tropical cirrus cloud and 30\% polar stratospheric cloud. The global cloud average represents a more accurate snapshot of Earth's atmosphere as it incorporates both the tropical and polar climates. The spectral resolution is kept to $0.02$~$\mu$m. The extreme synthetic cloud spectrum represents a global cloud layer from 15~km to 20~km with an aerosol extinction of 0.05. This extreme synthetic cloud has a higher extinction than real stratospheric clouds on Earth. Stratospheric clouds would have a greater impact on the spectra if they had a higher aerosol extinction e.g., due to greater volcanic activity or stratospheric humidity.}
    \label{fig:EffH}
\end{figure*}

In Figure~\ref{fig:EffH}, we show that tropical cirrus clouds increase the effective height for deep spectral windows. Some spectral features at effective heights below the cloud deck ($\sim$16~km) are still visible, because the clouds are not fully opaque. The peak features match, but are slightly higher than the world average non-cloudy spectrum. This is due to the lower transmission we see in the tropics in comparison to the world average, a result of higher atmospheric temperature and hence higher water vapour density present in these regions. 

We also see that polar stratospheric clouds reduce the ability to probe the lower layers of the atmosphere, and mute various peak features. This is due to the fact that polar stratospheric clouds have a greater transmission at higher altitudes than the world average non cloudy data. This can be attributed to colder temperatures present in the stratosphere with polar stratospheric clouds. Polar stratospheric clouds also contribute to ozone depletion in the atmosphere, thus the ozone feature at about $9.66$~$\mu$m may be further muted due to a lower ozone concentration~\citep{Zondolo}.  

\section{Observing Earth Analog Planets}
\label{exoplanet}
\subsection{Refraction}

We consider three planet-star systems: Earth--Sun, TRAPPIST-1e, and a hypothetical Earth orbitting the white dwarf, WD 1856+534. 

We choose the TRAPPIST-1 system because M-dwarf stars offer the best targets to search for life, in the near future, due to their small size and close-in habitable zones. We specifically consider TRAPPIST-1e because it has a similar bulk density to Earth and several groups have discussed its habitability with different atmospheric models~\citep{Krissansen-Totton2018,Lincowski2018,OMalley}. In general, the TRAPPIST-1 system is optimal for transit spectroscopy due to the large planet to star radius ratio~\citep{Gillon2017}. It is unlikely that TRAPPIST-1e will be an Earth twin because the increased stellar activity from its M-dwarf host could erode its atmosphere, but this could possibly be alleviated by degassing from the mantle~\citep{Moore2020}. Furthermore,~\cite{Wolf2017} and~\cite{Turbet2018} showed that TRAPPIST-1e could retain surface liquid water--and hence roughly Earth-like conditions--for a range of atmospheric compositions and thicknesses. Moreover, planets orbiting close to a late type M-dwarf star, like TRAPPIST-1e, will likely be tidally locked into synchronous rotation. The resulting climate is unlikely to support stratospheric clouds exactly like those on Earth. Lastly, for planets like TRAPPIST-1e, photochemical hazes may flatten out the transmission spectra more than condensate clouds~\citep{Fauchez2019}.

A hypothetical white dwarf (WD) system would offer an even better planet-to-star radius ratio since a WD is roughly the size of Earth. This would dramatically improve the signal to noise ratio for atmospheric characterization in comparison to other planetary systems. There are no known WD rocky worlds, but WD 1856+534 is known to host a Jupiter-like gas planet~\citep{Vanderburg2020} and countless ``polluted'' WDs attest to the prescense of rocky material in their vicinity~\citep{Doyle2021}. We assume a full transit with the WD system; grazing transits would reduce the overall transit depth.

Refraction stops the host star's light from probing the deeper layers of the planet's atmosphere, and thus creates a minimum effective thickness in the transit spectra. The refractivity increases with pressure because the angular deflection is proportional to the density of gas. The angular size of the host star in the planet's sky changes the range of angles at which the light probes the atmosphere. During a transit, the atmosphere is probed to a certain maximum pressure, $p_{\rm max}$, as the star's light will reach a critical deflection point within the atmosphere, given by~\citep{Kalt&Traub,Bet2014,Bet2015,Robinson}:

\begin{equation}
    \frac{p_{\rm max}}{p_0} = \frac{1}{v_0}\frac{R_p+R_*}{a}\sqrt{\frac{H}{2\pi R_p}},
\end{equation}
where $p_0$ is the surface pressure, $v_0$ is the refractivity of the atmosphere, and $a$ is the orbital distance of the planet. 

The minimum impact parameter, $b_{\rm min}$, at which the atmosphere can be probed is then

\begin{equation}
    b_{\rm min} = H\ln{\left(\frac{p_0}{p_{\rm max}}\right)}.
\end{equation}

Table~\ref{tab:bmin} lists the parameters used and the $b_{\rm min}$ for the three planet--star systems.

\begin{table}
\caption{The parameters for TRAPPIST-1e  were taken from ExoMast. The negative $b_{\rm{min}}$ indicates that photons can probe down to the surface of the planet and thus refraction does not interfere in this planet--star system; we take it as zero for the refraction calculations. The parameters for WD 1856+534 were taken from the 5000K WD in~\protect\cite{Kozakis}. The refractivity is $v_0=2.9\times10^{-4}$, and the scale height is $H=8.8$ km~\citep{Bet2015} to match Earth's atmosphere for all planet--star systems.}
\setlength{\tabcolsep}{0.1\tabcolsep}
\begin{tabular}{c|c|c|c|c}
\textbf{\begin{tabular}[c]{@{}c@{}}Planet--Star\\  System\end{tabular}} &
  \textbf{\begin{tabular}[c]{@{}c@{}}R$_{p}$\\ (R$_{E}$)\end{tabular}} &
  \textbf{\begin{tabular}[c]{@{}c@{}}R$_{*}$\\ (R$_{sol}$)\end{tabular}} &
  \textbf{\begin{tabular}[c]{@{}c@{}}a\\ (AU)\end{tabular}} &
  \textbf{\begin{tabular}[c]{@{}c@{}}b$_{\rm min}$\\ (km)\end{tabular}} \\ \hline
Earth and Sun                                                        & 1    & 1      & 1      & 12.6 \\ \hline
\begin{tabular}[c]{@{}c@{}}TRAPPIST-1e\end{tabular} & 0.91 & 0.1192 & 0.029  & -0.8 \\ \hline
\begin{tabular}[c]{@{}c@{}}Earth and \\ WD 1856+534\end{tabular}      & 1    & 0.0131 & 0.0096 & 5.2 
\end{tabular}
\label{tab:bmin}
\end{table}

\begin{figure*}
    \centering
    \includegraphics[scale=0.5]{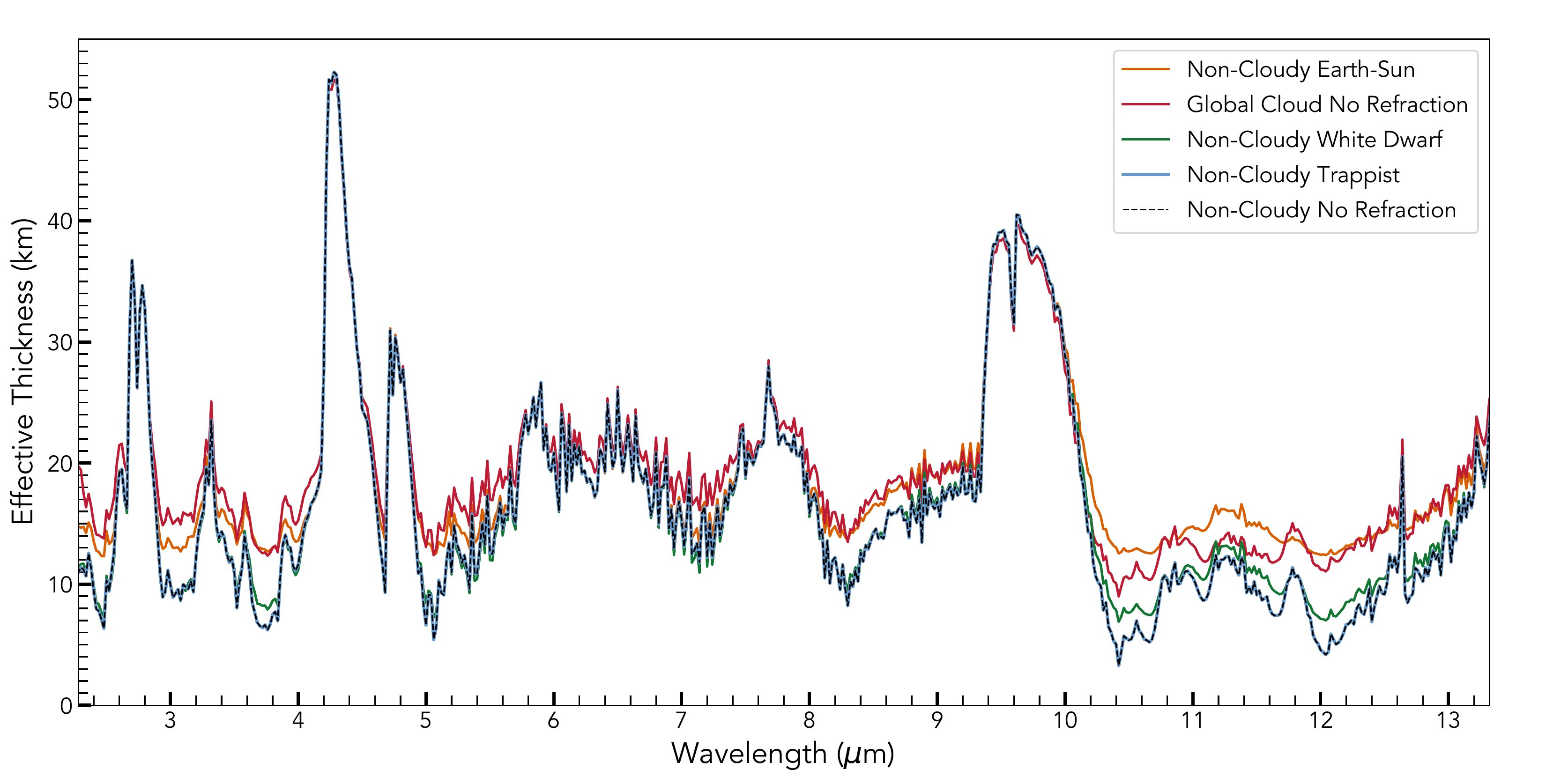}
    \caption{This figure shows the effects of refraction on the effective thickness spectra of hypothetical planets with an Earth-like atmosphere. TRAPPIST-1e has a negative $b_{\rm min}$, so transit spectroscopy of this planet is unaffected by refraction. As a result, it matches the dashed line which represents a non-cloudy planet where refraction is not taken into account. The ACE world average is taken for the non-cloudy atmospheres and the global cloud average atmosphere is taken for the cloudy atmosphere. The cloudy atmosphere is not affected by refraction in the WD system because the light does not probe down to $b_{\rm min}$ due to the clouds. }
    \label{fig:Refraction}
\end{figure*}

Figure~\ref{fig:Refraction} shows that planet--star systems where the star has a smaller angular size in the planet's sky are more vulnerable to atmospheric refraction. We can also see that the effects of clouds would overpower the effects of refraction for the TRAPPIST-1e and WD system, thus refraction need not be considered for these cases. Without accounting for noise, one could still distinguish between a cloudy and clear atmosphere for the TRAPPIST-1e and WD system but it would be difficult to do so for the Earth-Sun system. 

\subsection{Simulated JWST Observations}

To develop a realistic transit spectrum, we model JWST noise from NIRSpec and MIRI using PandExo~\citep{Pandexo}. For the TRAPPIST-1 system, the stellar and planetary parameters are taken from ExoMast~\citep{ExoMast}. TRAPPIST-1 is modelled with a temperature of $2559$ K, metallicity of  [Fe/H]$=0.04$, surface gravity of $\log(g) = 5.28$, and a J band magnitude of $11.354$. The transit duration for TRAPPIST-1e is $0.0397$ days. We assume equal amounts of observing time in transit as out of transit~\citep{LustigYaeger}. 

For the WD system, we adopt the following parameters from~\cite{Kaltenegger/WD}. For WD 1856+534, we assume a temperature of $4780$ K, J band magnitude of $15.677$, and metallicity of 0.005. The surface gravity of a WD would be larger than $\log(g)=5.5$, however this was the upper limit set for this parameter on Pandexo and should not significantly affect the SNR estimate. The planetary parameters were set to match Earth. A transit time of $2.2$ min and a total observing time of $1.5$ hr is used for the noise model. 

Two separate PHOENIX stellar models are used to simulate the photosphere for the TRAPPIST-1e and WD systems~\citep{Pheonix}. To model Earth-analog planets, we use the ACE-derived spectra of non-cloudy and cloudy Earth-like atmospheres for the planetary models. The non-cloudy models take into account the effects of refraction as needed. We assume a saturation limit of $80\%$ full well. We consider two JWST instrument modes: the g395m disperser with R=1000 is used for NIRSpec, and the Slitless mode is used from MIRI. We adopt noise floors of $75$~ppm for NIRSpec~\citep{Ferruit}, and of $40$~ppm for MIRI, since~\cite{Greene} and~\citep{Beichman} report noise floors of $50$~ppm and $30$~ppm respectively.

\begin{figure*}
    \centering
    \includegraphics[scale=0.45]{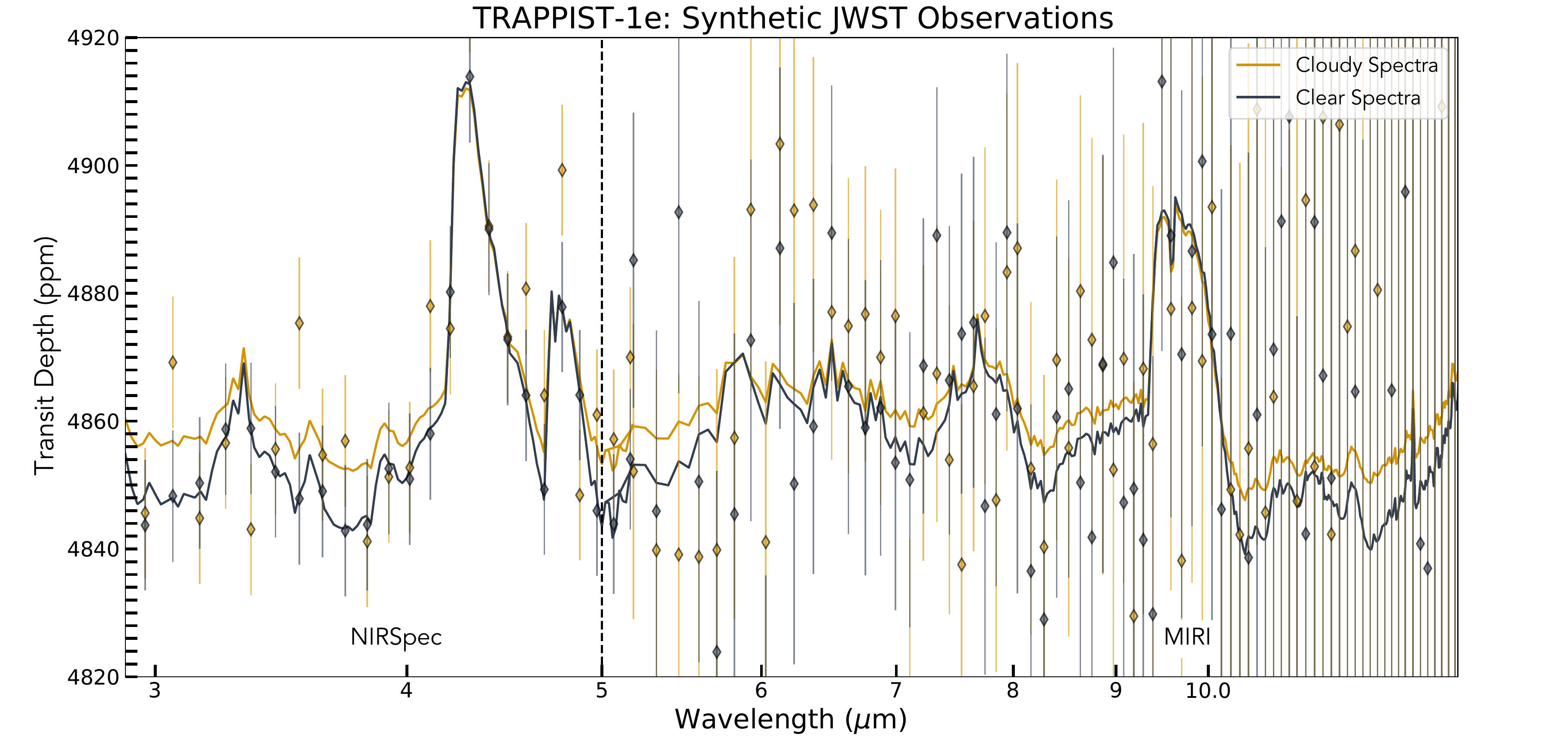}
    \caption{Pandexo simulations showing synthetic MIRI and NIRSpec observations for a cloudy and clear Earth-like atmosphere in the TRAPPIST-1e system. The data represents the combined data of 150 transits and are binned to $0.1$ $\mu\mathrm{m}$. It would be difficult to distinguish between the cloudy and clear scenarios based on such data thus, transit spectroscopy would be largely unaffected by clouds.}
    \label{fig:Trappist}
\end{figure*}

\section{Atmospheric Characterization}
\label{characterization}
\subsection{Distinguishing between a Cloudy and Clear Stratosphere}

To determine the number of transits needed to distinguish between a clear and cloudy atmosphere for the TRAPPIST--1 and WD systems, we compute the normalized root-mean squared residuals, NRMSR, following~\cite{LustigYaeger}. Instead of comparing transit observations to a featureless spectra, we treat the clear spectra as the baseline and quantify the effect of clouds. The NRMSR only depends on the difference between the cloudy and clear model, and the instrumental uncertainty predicted by Pandexo. The NRMSR is calculated for the entire spectral range, $N_{\lambda}$, using 

\begin{equation}
    \langle {\rm NRMSR} \rangle = \sqrt{\sum_{i=1}^{N_{\lambda}}\left(\frac{{\rm cloud}_{i}-{\rm clear}_{i}}{\sigma_i}\right)^2},
\end{equation}
where ${\rm cloud}_{i}$ and ${\rm clear}_{i}$ are the cloudy and clear transit spectra, and $\sigma_i$ is the uncertainty associated with $\lambda_i$. 

\begin{figure*}
    \centering
    \includegraphics[scale=0.45]{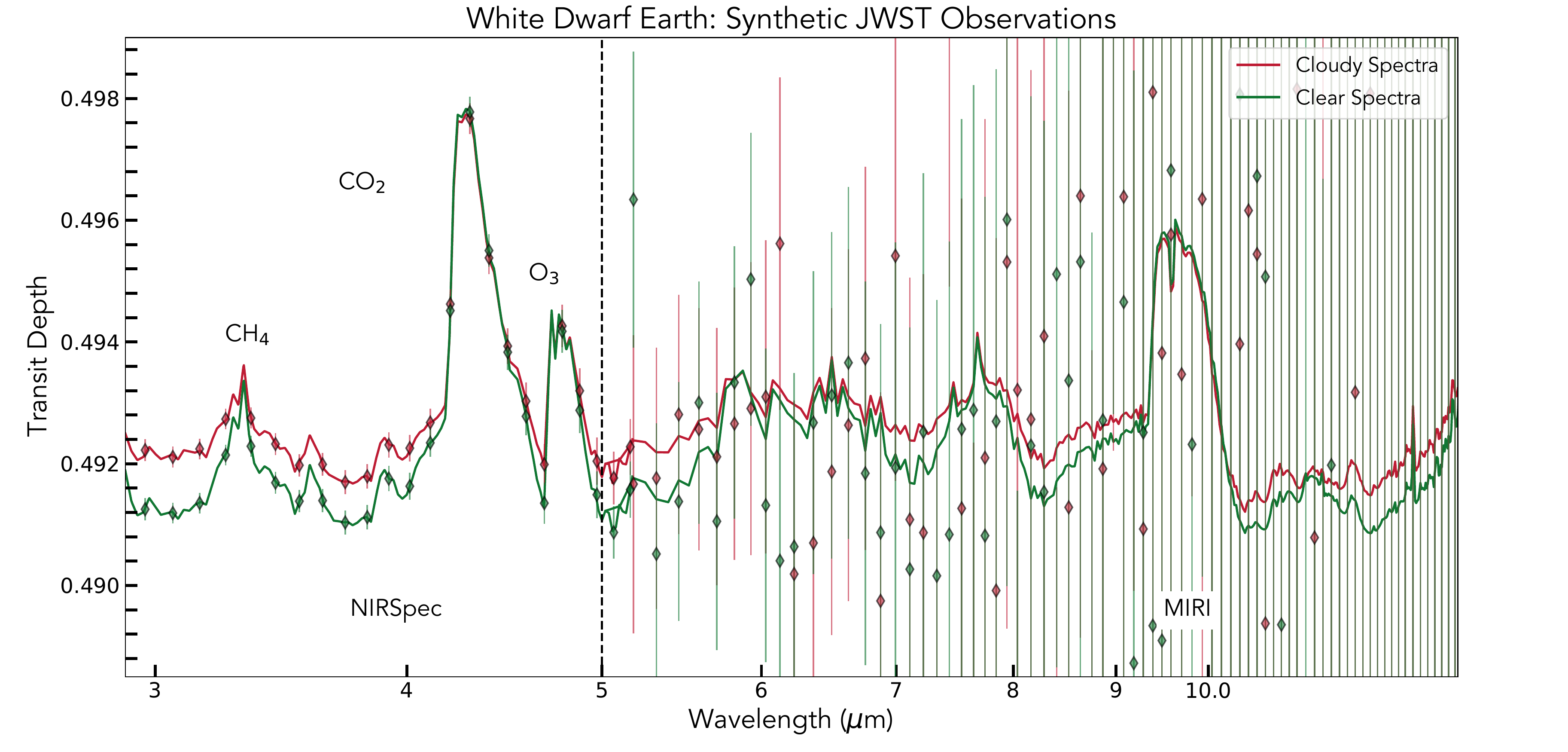}
    \caption{Pandexo simulations of NIRSpec and MIRI transit spectra of an Earth-analog planet orbiting WD 1856+534. The NIRSpec data represents four transits, unlike the MIRI portion which represents 150 transits; the data is binned to $0.1$ $\mu\mathrm{m}$.}
    \label{fig:WD}
\end{figure*}

After 150 transits, neither MIRI or NIRSpec will distinguish the clouds and clear scenarios at NRMSR of 10 (Figure~\ref{fig:Trappist}). The maximum difference seen between the cloudy and clear spectra is approximately $10-15$ ppm. A noise floor smaller than this difference would be required to distinguish between the two cases, which is much lower than the assumed noise floor for either instrument. As a result, the relatively pessimistic noise floor assumptions do not bias the result. In other words, even widespread stratospheric clouds would not be detectable on TRAPPIST-1e if it had an Earth-like atmosphere. For the WD system, MIRI observations could not reach an NRMSR of 10 after 150 transits, but an NRMSR of 10 would be achieved after 4 transits using NIRSpec (Figure~\ref{fig:WD}). 

\subsection{Spectral Retrievals} 

We test whether spectral retrieval can distinguish between a cloudy and clear atmosphere for TRAPPIST-1e. Spectral retrieval is a commonly used tool to interpret exoplanetary transit spectra~\citep{Madhusudhan2018}. Retrievals will generate millions of spectra for a wide range of parameters using Bayesian sampling algorithms to find the parameters that best match the observations. Our retrievals are performed with TauREx 3.0~\citep{Taurex}. 

The TauREx forward model uses a 6-point temperature profile modelled on the spring-fall pressure-temperature profile from~\cite{COESA}. The atmosphere is divided into 100 uniformly spaced layers in a log grid, ranging from $10^{5}\textup{--}10^{-2}$ Pa. The atmosphere is N$_2$ and O$_2$, with the following spectroscopically active gases: CO$_2$, H$_2$O, CH$_4$, and O$_3$. The molecular cross sections were taken from ExoTransmit~\citep{ExoTransmit}. The forward model takes into account the effects of absorption, collision-induced absorption (CIA), and Rayleigh scattering. HITRAN~\citep{Hitran} CIA data are used for the various molecule-molecule interactions: N$_2$-N$_2$, O$_2$-O$_2$, O$_2$-N$_2$, N$_2$-H$_2$O, O$_2$-CO$_2$, CO$_2$-CO$_2$, and CO$_2$-CH$_4$. The volume mixing ratios of molecules can vary with altitude, but we consider vertically uniform abundances. Methane and carbon dioxide abundances are vertically uniform in Earth's atmosphere, but ozone and water vary with the pressure; specifically, water vapour is less abundant above the cloud deck. As a result, there may be a discrepancy between our retrieved abundances for these molecules and their true abundances as well as predictions made from GCMs. For our observed spectrum, we use the Pandexo noise model for the TRAPPIST-1e clear and cloudy atmospheres with 150 transits using NIRSpec. The retrieval is conducted using the Nestle Optimizer package, where the mixing ratios of the active gases, planetary radius, and altitude of the cloud deck are free parameters.

\begin{figure*}
    \centering
    \includegraphics[scale=0.45]{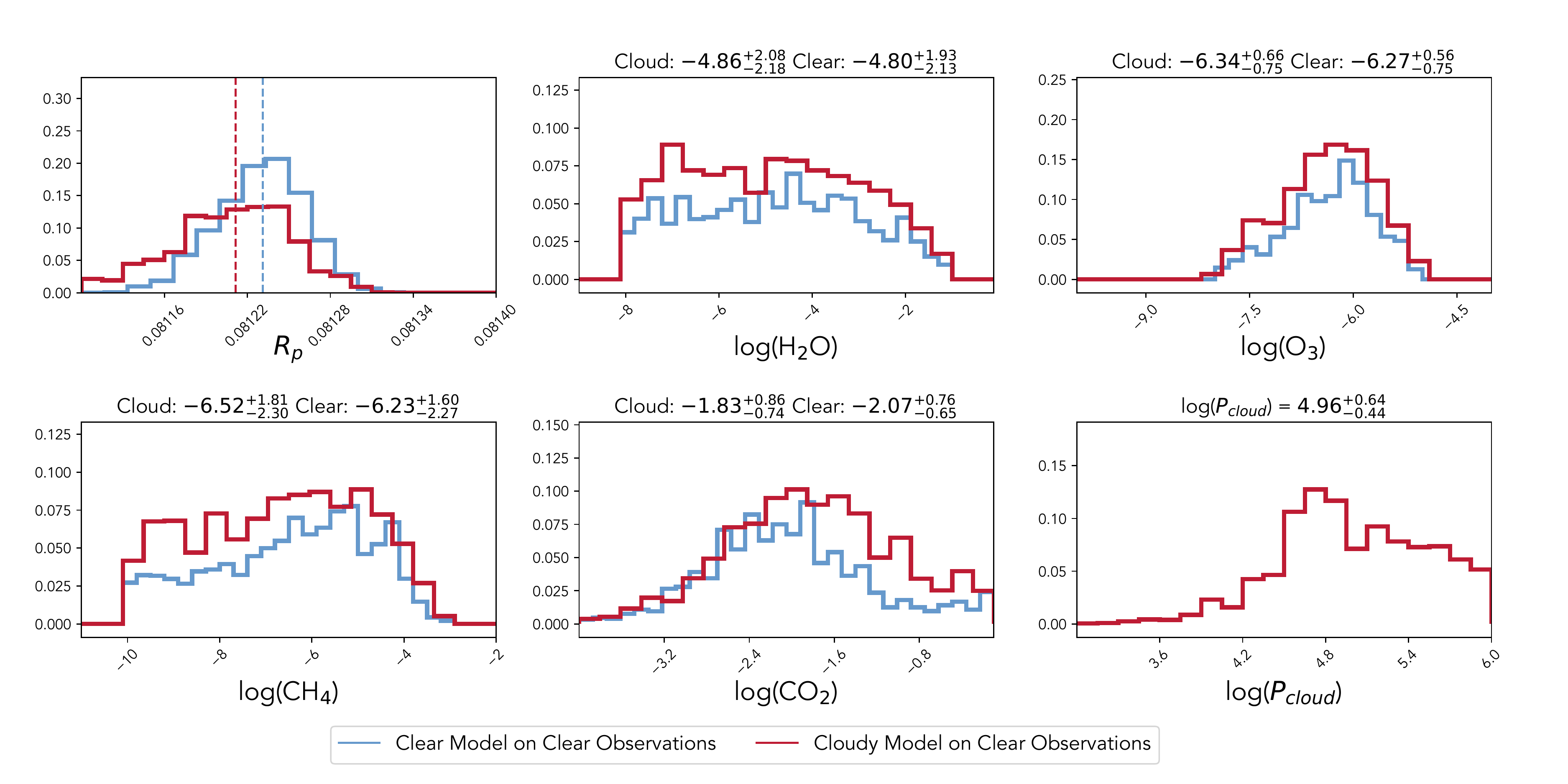}
    \includegraphics[scale=0.45]{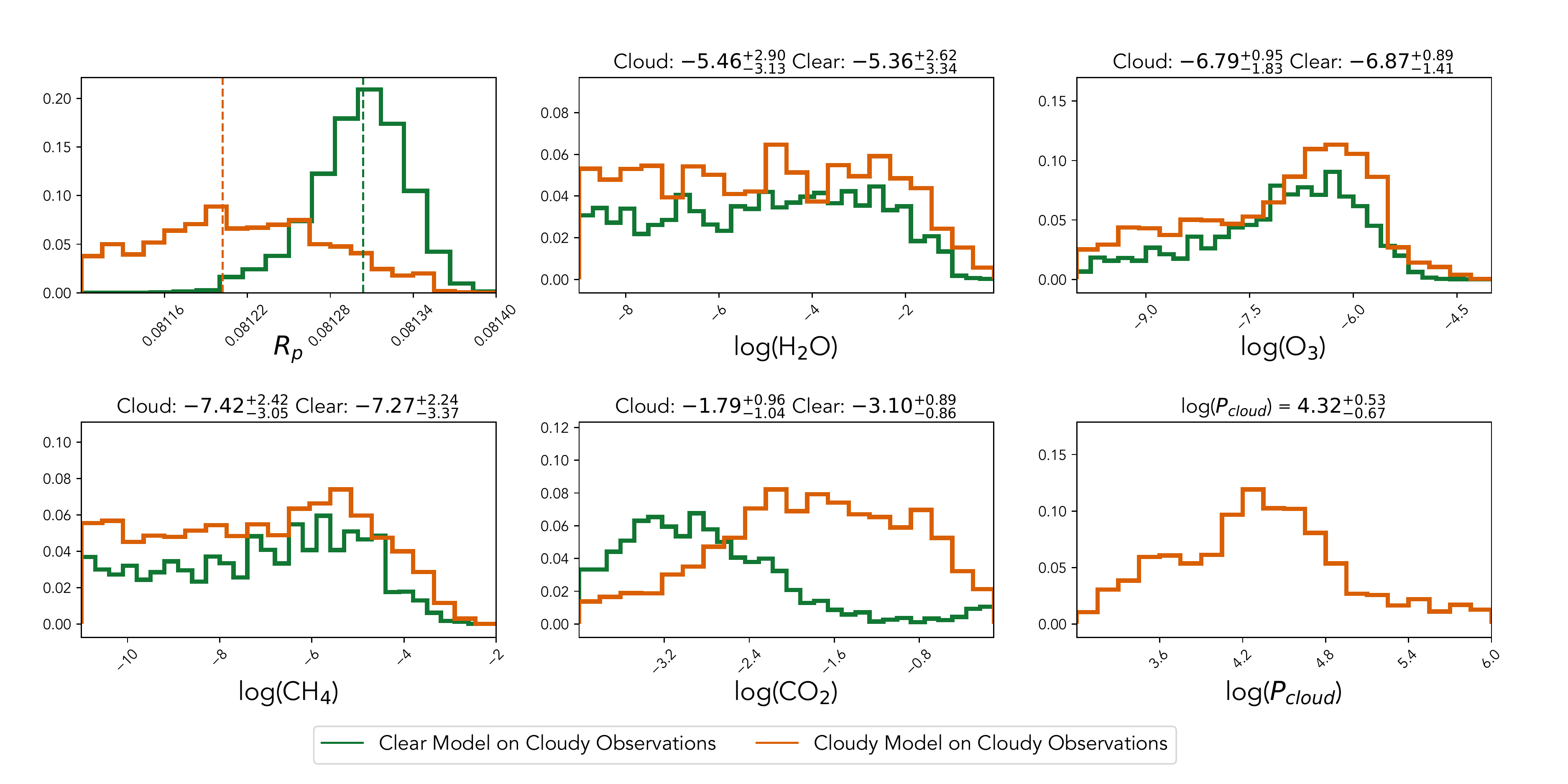}
    \caption{The retrievals for four different situations for 150 transits of TRAPPIST-1e using NIRSpec.  The vertical dashed lines represent the maximum a posteriori values for the planetary radius in terms of Jupiter radii. The log of the molecular abundance and the log of the pressure at the cloud deck are presented as calculated by TauRex3.0. We provide the 1$\sigma$ confidence interval for each retrieval choice. }
    \label{fig:Histo}
\end{figure*}

We attempt retrievals with two models, one without clouds, and one with a completely opaque global grey cloud layer. In Figure~\ref{fig:Histo} we present the retrievals in which we compare both cloudy and clear JWST TRAPPIST-1e observations to the two different models. The Bayesian Information Criterion, BIC, is used for model selection, where a lower value is preferred~\citep{Schwarz1978,Raftery1995}. Table~\ref{tab:BIC} displays the BIC values calculated for the four different scenarios to evaluate which model is preferred. The $\Delta$BIC value is less than 3 for both the cloudy and clear observations. We find that there is no strongly preferred model for a clear or cloudy exo-Earth in the TRAPPIST system. This indicates that spectral retrieval performed on JWST observations cannot strongly detect or rule out stratospheric clouds on an Earth-like TRAPPIST-1e. However, other missions such as the Origins Space Telescope~\citep{Leisawitz2021} may offer a better signal to noise ratio and thus could detect these types of clouds.

\setlength{\tabcolsep}{0.47\tabcolsep}
\begin{table}
\caption{The BIC values for each of the four difference scenarios. The preferred model for the cloudy observations is the one without a cloud in the retrieved model as it has a lower BIC value. However, the $\Delta$BIC for the cloudy observations is below 1, thus there is not statistically stronger model. While $\Delta$BIC for the clear observations is above 2, making it a positive detection, this still does not qualify as a strong detection.}
\begin{tabular}{|c|c|c|c|}
\hline
\multirow{2}{*}{Observations} &
  \multirow{2}{*}{\begin{tabular}[c]{@{}c@{}}Cloud in \\ Retrieved Model\end{tabular}} &
  \multirow{2}{*}{\begin{tabular}[c]{@{}c@{}}No Cloud in\\  Retrieved Model\end{tabular}} &
  \multirow{2}{*}{$\Delta$BIC} \\
       &       &       &      \\ \hline
Cloudy & 26.55 & 25.63 & 0.92 \\ \hline
Clear  & 28.96 & 26.03 & 2.93 \\ \hline
\end{tabular}
\label{tab:BIC}
\end{table}

The posteriors from a cloudy retrieval of cloudy observations shown in Figure~\ref{fig:Histo} provide lower abundances than those from clear observations for H$_2$O, CH$_4$, and O$_3$. As one might expect, the posteriors are wider in the cloudy case than the clear case, indicating that there is higher uncertainty in the measured abundances with cloudy observations. Our retrievals find a higher abundance for CO$_2$ for the cloudy case, however the uncertainty is larger as well. \cite{Krissansen-Totton2018} find that their cloudy posteriors are also wider than their clear posteriors for an Archean Earth TRAPPIST-1e. Their retrievals for the abundance of  H$_2$O and CH$_4$ are also lower for their cloudy observations in comparison to their clear observations. Moreover, the retrieved altitude for the cloud deck agrees with the true value from our observational data, given the uncertainties.

If we assume the retrieval with clear observations offers the best estimate for the planetary radius, we find that the retrieval code underestimates the planetary radius for cloudy observations with a cloud in the model, and significantly overestimates the planetary radius for cloudy observations with a clear model. Moreover, there is a large discrepancy in the retrieval for the abundance of CO$_2$ for the cloudy observations depending on whether one uses the cloudy or non-cloudy retrieval; clouds are needed in the retrieval in order to obtain accurate constraints on CO$_2$. However, as there is no strong preferred model for the cloudy observations, the planetary radius and molecular abundance of CO$_2$ would remain uncertain.

\section{Discussion \& Conclusions}
\label{discussion}

A limiting factor for transit spectroscopy is the presence of high-altitude aerosols in the form of photochemical hazes or condensate clouds. We have focused on the latter, which dominate modern Earth and most simulations of temperate terrestrial exoplanets.

We now compare our results to other studies of transit spectroscopy for cloudy terrestrial planets. \cite{Mayorga} modelled cirrus clouds at 8.5 km altitude with an optical depth of 3 and showed that these clouds increase the altitude down to which the atmosphere can be probed in transit. Meanwhile, our stratospheric clouds have a higher cloud deck but lower optical depth and do not greatly affect the transit spectrum. \cite{Mayorga} note that solar occultation data will downplay the effects of refraction on spectra. However, a more comprehensive approach to adding the effects of refraction to our data would lessen the difference between the clear and cloudy spectra, making it more difficult to differentiate between the two. 

GCMs of TRAPPIST-1 planets with Earth-like atmospheres suggest that clouds would be the single limiting factor in characterizing the atmosphere. \cite{Komacek} developed an ExoCam GCM of an aqua-planet consisting of only N$_2$ and H$_2$O orbitting an M-dwarf star. With their model, clouds increase the number of transits required to detect water features with JWST by a factor of 10--100, and their transit features differed by up to 20 ppm for clear and cloudy atmospheres. Similarly, ~\cite{Suissa2020} also model water rich Earth sized planets using Exocam and find that clouds dominate the spectral features. Our Earth-inspired stratospheric clouds are optically thin, even in transit, leading to their smaller impact on the transit spectrum. This is due to the fact that Earth's stratosphere is dry. Meanwhile,~\cite{Komacek} and~\cite{Suissa2020} use an M-dwarf spectrum to model their exoplanetary atmospheres, which would lead to higher temperatures and humidity in the stratospheres \citep{Fujii2017}, likely resulting in more optically thick clouds. Moreover, in modelling an aqua-planet without any continents, this allows for a higher water vapor content leading to enhanced cloud formation~\citep{Lewis2018}. As a result, the clouds generated by ExoCam for close-in tidally locked rocky planets with aqua-planet surfaces would be optically thicker and have a larger impact on the transit spectra than the ones seen on Earth.

Similarly, \cite{Pidhorodetska2020} model a modern Earth atmosphere on TRAPPIST-1e with a global deep ocean. They use a GCM developped by the Laboratoire de Météorologie Dynamique (LMD) that produces clouds at about 15~km that are completely opaque to infrared and visible radiation. The transit features between their clear modern Earth and cloudy modern Earth differed by about 15--20~ppm, a larger difference than we found. While their cloud deck is lower than our stratospheric clouds, ours are not fully opaque and thus layers below the cloud deck can still be probed. 

In comparison to other GCM models, ROCKE-3D produces thinner clouds. ~\cite{Fujii2017}, use ROCKE-3D to simulate Earth sized aqua-planets orbiting the red dwarf GJ 876. Their GCM produces optically thin clouds at high altitudes around the terminator for a solar incident flux similar to the Earth's.  Generally, their clear spectrum is similar to the one presented in this work, as the lowest effective thickness reaches approximately 8--9 km in both cases. The clouds produced by their GCM around the terminator are similar to our stratospheric clouds as they are optically thin and at a higher altitude; we see a maximum increase of 8.5 km in the effective thickness of Earth's atmosphere from clouds whereas they see a difference of about 10 km. These results are fairly similar and we can attribute the small differences to slight changes in the atmospheric makeup of the two models. Greater abundance of CO$_2$ would result in warmer climates and thus an enhanced water vapour mixing ratio and more, higher altitude clouds~\citep{Wolf2017, May2021}. ~\cite{Fujii2017} use a much lower concentration of pCO$_2$, about 1 ppm, but still have a higher effective thickness than the one observed in this paper.

Inspired by the work of~\cite{Krissansen-Totton2018},~\cite{Mikal-Evans} uses a Bayesian evidence framework to determine the confidence level at which CH$_4$ and CO$_2$ can be detected in an Archean Earth atmosphere with the presence of clouds and/or photochemical hazes. They find that a 5$\sigma$ detection of both CH$_4$ and CO$_2$ can be made with only 5--10 co-added transits with clouds at 600--100 mbar or about 16 km above the surface. The same strong detection requires more co-added transits as the cloud deck is placed higher in the atmosphere. This matches our results, as our cloud deck is at approximately 17 km, and we find that our clouds would not impede the detectability of bio-signatures such as CH$_4$ and CO$_2$.

The small change in effective thickness reported here means we are unable to detect stratospheric clouds on a temperate, terrestrial planet orbitting a M-dwarf if it has an Earth-like atmosphere. This could have other implications in terms of understanding the planet's characteristics. While our work focused on tropical cirrus clouds and polar stratospheric clouds, the effect of mild volcanic stratospheric clouds would have similar results and would be ultimately undetectable~\citep{Kaltenegger2010}. These types of clouds have been used in many ``geoengineering'' proposals to artificially alter atmospheric abundances and control the planet's climate~\citep{,Keith2016,Cziczo2019}. As a result, we would not be able to identify these potential artifacts of extraterrestrial intelligent life.  

In summary, we used clear and cloudy solar occulation data to create synthetic transit spectra for the TRAPPIST-1e and a hypothetical habitable WD system. We found that the effect of Earth-like stratospheric clouds overpowers the effects of refraction for these planetary systems. Moreover, the clear and cloudy spectra do not differ greatly, as the maximum difference in effective thickness is 8.5 km at 2.28 $\mu$m. JWST could detect stratospheric clouds on a hypothetical Earth twin orbitting a white dwarf system. However, due to instrumental noise, even with 150 transits JWST would not be able to significantly detect or rule out the presence of these clouds on TRAPPIST-1e if it hosted an Earth-like atmosphere. JWST's ability to detect these clouds would decrease with a more realistic number of transits, therefore it's highly unlikely that JWST would have the capability to find stratospheric clouds on TRAPPIST-1e if it's atmosphere was exactly like Earth. This implies that Earth-like stratospheric clouds should not significantly impact the number of transits needed to detect bio-signatures.

\section{Acknowledgements}
This work was funded by a Trottier Excellence Grant and a Technologies for Exo-Planetary Science (TEPS) Undergraduate Fellowship. The authors would like to thank Chris Boone and Peter Bernath for providing the cloudy ACE-FTS data. This work was supported by the Institute for Research on Exoplanets (iREx), McGill Space Institute (MSI), and McGill Exoplanet Characterization Alliance (MEChA). We would also like to thank the anonymous referee for their incredibly helpful and detailed comments which greatly improved this work. 

\section*{Data Availability}
The data underlying this article will be shared on reasonable request to the corresponding author.

\appendix
\section{Occultation Data}
\label{A1}
\begin{table*}
\caption{Details of the specific SCISAT ACE-FTS solar occultations used in this work. The cloud type refers to which cloud was present during the occultation measurement at the specific date and location. The beta angle dictates the vertical range of measurements taken during the solar occultation. }
\begin{tabular}{c|c|c|c|c}
\textbf{\begin{tabular}[c]{@{}c@{}}Occultation\\  Name\end{tabular}} &
  \textbf{\begin{tabular}[c]{@{}c@{}}Cloud \\ Type\end{tabular}} &
  \textbf{Coordinates} &
  \textbf{\begin{tabular}[c]{@{}c@{}}Date\\ (yyyy-mm-dd)\end{tabular}} &
  \textbf{\begin{tabular}[c]{@{}c@{}}Beta \\ Angle ($\degree$)\end{tabular}} \\ \hline
sr79241 & tropical cirrus & (-1.12,99.55)   & 2018-04-28 & -53.11 \\ \hline
sr79236 & tropical cirrus & (-3.40,-137.84) & 2018-04-28 & -53.75 \\ \hline
ss83526 & tropical cirrus & (12.58,-126.47) & 2019-02-13 & 56.36  \\ \hline
ss11637 & tropical cirrus & (8.93,48.15)    & 2005-10-10 & 59.61  \\ \hline
sr77903 & polar stratospheric  & (66.02,29.86)   & 2018-01-28 & 18.87  \\ \hline
sr7874  & polar stratospheric  & (65.47,25.36)   & 2005-01-28 & 32.56 
\end{tabular}

\label{tab:occultations}
\end{table*}

\bibliography{references}

\bsp	
\label{lastpage}
\end{document}